\documentclass[a4paper,12pt]{article}
\usepackage[cp1251]{inputenc}
\usepackage[dvips]{graphicx}
\graphicspath{{.}}

\begin{document}
\mathsurround=2pt \sloppy
%\begin{center}
\title{\bf Low Temperature Limit of Stability of Coherent Precession of Spin in the Superfluid $^{3}$He-B}
\author{E.V. Surovtsev,  I.A. Fomin \\{\it P.L.Kapitza Institute for Physical Problems}
\\ {\it Moscow Russia}}
%\end{center}
\maketitle
\begin{abstract}
It is shown that instability of homogeneous precession is caused
by combined effect of anisotropy of spin wave velocities and
dipole interaction. In the principal order on the ratio of the
Leggett frequency to the Larmor frequency  the increments of
growth of spin wave amplitudes are found. The magnitude of the
maximum increment for all deviation angles of spin from its
equilibrium orientation is calculated. The estimation is made of
the minimum temperature down to which the precession is stable.
\end{abstract}

{\bf 1}. In a contrast to the superfluid A-phase, in the B-phase
long wavelength
 perturbations do not destroy homogeneous precession of
 magnetization.
Precession is stable with respect to such perturbation if the
initial deviation angle of magnetization $\beta>\theta_0$, where
$\theta_0=\arccos(-1/4)$ and marginally stable if
$\beta\leq\theta_0$. However, at low temperatures fast decay of
homogeneous precession is observed in both cases. Such decay was
first observed in experiments \cite{nyeki,Bunkov_precession} and is
referred as {\it catastrophic relaxation}. Earlier we considered a
process of parametric excitation of spin waves with finite wave
vectors by the precession of magnetization as a possible origin of
the decay \cite{sf}. In the theory of magnetics this effect is known
as the {\it Suhl instability} \cite{suhl}. The distinctive property
of the instability in $^3He$-B is that it occurs at the precession
with large tipping angles $\beta\sim 100^o$ and also that excitation
of different types of spin waves is possible. In Ref. \cite{sf} a
scheme was proposed that takes into account both mentioned
peculiarities and increments of instability for each of three types
of spin waves were found. The dependence of the obtained increments
on magnetic field does not agree with that experimentally observed
\cite{Lee}. The magnitudes of the increments were also overvalued.
The disagreement with the experiment was caused by our technical
mistake. As a result of this mistake it turned out, that the
instability of the precession can be induced by anisotropy of spin
wave velocities in $^3$He-B alone. The further analysis has shown
that in order to provide coupling between spin waves and precession
the dipole interaction has to be taken into account. It turned out
also that considerable contribution to the instability comes from
joint resonances when spin waves belonging to different branches of
spectra are excited simultaneously.

In the present paper a revised version of the theory of parametric
instability of homogeneous precession of spin in $^3$He-B is
presented. At the same time the foregoing disagreement with the
experiment is resolved. A comparison is also made with the results
of Ref.\cite{blv1,blv2}, where contribution of the boundaries to
the development of parametric instability is considered.

{\bf 2}. Following the procedure of Ref.\cite{sf} let us
parameterize orientation of the order parameter of $^3$He-B, which
is a rotation matrix  $R_{\xi i}$, by the Euler angles $\alpha$,
$\beta$, $\gamma$ ($z$-axis is oriented opposite to the direction
of the d.c. magnetic field {$\bf H_{0}$}). Actually, it is more
convenient to use the sum $\Phi=\alpha+\gamma$ instead of the
angle $\gamma$. Canonically conjugated momenta to these
coordinates are correspondingly the following combinations of spin
projections $P=S_{z}-S_{\zeta}$, $S_{\beta}$, $S_{\zeta}$, where
$S_{z}$
--- is projection of spin onto \emph{z}-axis, $S_{\zeta}$ --- its
projection onto $\zeta=\hat{R}\hat{\emph{z}}$ and $S_{\beta}$
--- is projection on the line of nodes (see for example \cite{Fomin1}).
Equations of motion are hamiltonian with respect to the stated
pairs of variables with the Hamiltonian
  \begin{equation}
     \label{Hamiltonian}
     H=\frac{1}{1+\cos{\beta}}\{S_{\zeta}^2+P
     S_{\zeta}+\frac{P^2}{2(1-\cos{\beta})}\}+\frac{1}{2}S_{\beta}^2+F_{\nabla}-\omega_{L}(P+S_{\zeta})+
     U_{D}(\alpha,\beta,\Phi).
  \end{equation}
 Here $\omega_L$ --- is the Larmor frequency, corresponding to the d.c.
 magnetic field,$F_{\nabla}$ --- the gradient energy,
 $U_{D}(\alpha,\beta,\Phi)$ --- the  dipole energy, which is of the order of
 the squared Leggett frequency in the $^3He$-B: $\Omega_B^2$.
We choose the units of measurements so that the magnetic
susceptibility of $^{3}He-B$ --- $\chi$, and the gyromagnetic
ratio for nuclei of $^{3}He$
--- \emph{g} are equal to unity. In this units spin has a dimensionality of
frequency and energy --- of a squared frequency correspondingly.
In the standard setting of NMR experiments in the regions distant
from the walls of the cell spin precesses in the so called
\emph{Leggett configuration}, when the "orbital vector"
$l_i=-R_{\xi i}s_{\xi}$ \cite{blv1} is parallel to the magnetic
field. In this case $U_{D}(\alpha,\beta,\Phi)$ does not depend on
the angle $\alpha$ and precession is described by the stationary
solution of equations of spin dynamics which do not contain
oscillating terms:
 \begin{eqnarray}
  \label{stat_solution}
   \alpha=-\omega_p t+\alpha_0,{~~}\gamma=\omega_p t+\Phi^{(0)}-\alpha_0,\nonumber\\
   P^{(0)}=\omega_p(\cos\beta-1),{~~}S_\beta^{(0)}=0,{~~}S_z^{(0)}=\omega_p\cos\beta,
 \end{eqnarray}
 where $\omega_p$ -- the frequency of precession.
 If $\beta>\theta_0$ then $\Phi^{(0)}=0$, and if $\beta<\theta_0$ --
\begin{equation}
\label{phi}
\cos\Phi^{(0)}=(\frac{1}{2}-\cos\beta^{(0)})/(1+\cos\beta^{(0)}).
\end{equation}
Explicit time dependence of the stationary solution
(\ref{stat_solution}) can be excluded if one transfers to the
variable  $\psi=\alpha+\omega_pt$ and uses new Hamiltonian
$\tilde{H}=H+\omega_pP$, then $\frac{\partial\psi}{\partial t}=0$.
To find  the spectra of excitation against the background of
precession we linearize the equations of motion on small
deviations from the stationary solution (\ref{stat_solution},
\ref{phi}): $\delta\psi(\textbf{r},t)=\psi-\psi^{(0)}$, etc. For
the sake of convenience  the following combinations of the
mentioned deviations are used:
\begin{eqnarray}
     \label{local coord}
\nu=\sin\beta\delta\psi, ~~ \vartheta=\frac{\delta P+(1-\cos\beta)\delta S_{\zeta}}{\omega_L\sin\beta},\nonumber \\
\varepsilon=\delta\Phi-(1-\cos\beta)\delta\psi,~~\sigma=\delta S_{\zeta}/\omega_L,~~\\
\zeta=\delta
S_{\beta}/\omega_L,~~\eta=-\delta\beta.{~~~~~~~~~~~~~~~}\nonumber
\end{eqnarray}
Expression for the gradient energy of $^{3}He-B$
 contains two coefficients that can be written as velocities of two
 types of spin waves $c_{\|}$ and
 $c_{\bot}$.
 In what follows units of length and time are chosen so that $\omega_L=1$ and $c_{\|}=1$.
 Without loss of generality one can assume that variables change only in $y$ and $z$ directions,
  then the time-independent part of the
  gradient energy has the form:
\begin{equation}
      \label{F_st}
       F_{\nabla
      st}= \frac{1}{2} [(1-\mu)(\nu_{,y}^2+\eta_{,y}^2)+\nu_{,z}^2+\eta_{,z}^2+\varepsilon_{,y}^2+(1-2\mu)\varepsilon_{,z}^2],
\end{equation}
where $\mu=1-c_{\bot}^2/c_{\|}^2$ -- is the anisotropy of spin
wave velocities. Parameter $\mu$ will be considered as a small
one, in fact $\mu\approx$1/4 \cite{dm_osc}. Furthermore,
$F_{\nabla}$ also contains term oscillating with the frequency of
precession:
\begin{equation}
      \label{F_oscill1}
       F_{\nabla
      osc1}= -\mu[(\eta_{,y}\varepsilon_{,z}+\eta_{,z}\varepsilon_{,y}) \cos\omega_p t+
      (\nu_{,y}\varepsilon_{,z}+\nu_{,z}\varepsilon_{,y})\sin\omega_p t]
 \end{equation}
and with the doubled frequency of precession
 \begin{equation}
      \label{F_oscill2}
 F_{\nabla
      osc2}= -\frac{\mu}{2}[(\eta_{,y}^2-\nu_{,y}^2)\cos{2\omega_p
      t}+2\nu_{,y}\eta_{,y}\sin{2\omega_pt}].
 \end{equation}
 In zero order approximation on small parameters $\mu$ and  $(\Omega_B/\omega_L)^2$ the equations for deviations
 have hamiltonian form with the Hamiltonian:
\begin{equation}
\label{h_1}
      h=\frac{1}{2}[(\vartheta+\eta)^2+\zeta^2+\sigma^2+ (\nabla\nu)^2+(\nabla\eta)^2+(\nabla\varepsilon)^2]
\end{equation}
and with respect to the pairs of canonically conjugated variables
$(\varepsilon,\sigma)$; $(\nu,\vartheta)$; $(\zeta,\eta)$. In each
pair the first variable is coordinate and the second is momentum.
Equations of motion for the pair $(\varepsilon,\sigma)$ have the
form:
 \begin{eqnarray}
     \label{longitud}
     \frac{\partial\varepsilon}{\partial t}=\sigma,{~~~} \frac{\partial\sigma}{\partial t}=\Delta\varepsilon.{~~}
  \end{eqnarray}
It is convenient to rewrite them in a vectorial form:
\begin{equation}
          \label{X_equation}
          \frac{d\mathbf{X}_1}{dt}=\hat{M}_1\mathbf{X}_1.%\frac{d\mathbf{X}}{dt}=\hat{M}_0\mathbf{X}
    \end{equation}
 Solutions of the system
(\ref{longitud}) have the form of plane waves
$\textbf{e}^{\pm}_1\exp[i(\textbf{k}\textbf{r}\mp\omega_1 t)]$
with the dispersion law

    \begin{equation}
    \label{fr-long}
        \omega_1=k.
         \end{equation}
Here $\textbf{e}^{\pm}_1$ are right eigenvectors of matrix
$\hat{M}_1$ corresponding to eigenvalues  $\mp i\omega_1$:
$$
\textbf{e}^{\pm}_1=\left(
\begin{array}{c}
1\\
\mp i\omega_1
\end{array}
\right). \eqno \stepcounter{equation}(\arabic{equation})
$$
One needs left eigenvectors $\textbf{f}^{\pm}_1$ of the same
matrix to be able to make projections. They can be normalized so
that the following conditions are met:
 \begin{equation}
    \label{covar1}
 (\textbf{f}^+_1,\textbf{e}^+_1)=1;{~~}
 (\textbf{f}^+_1,\textbf{e}^-_1)=0.
         \end{equation}
Here the scalar product is defined as
\begin{equation}
    \label{dot}
 (\textbf{a},\textbf{b})=\sum_{n=1,2}a_n^*b_n.
         \end{equation}
As a result
\begin{equation}
    \label{covar2}
\textbf{f}^{\pm}_1=\frac{1}{2\omega_1}\left(\omega_1; \mp
i\right).
\end{equation}
In a similar way the following dispersion laws for two transverse
modes are obtained:
 \begin{equation}
    \label{tr2}
    \omega_2=\sqrt{\frac{1}{4}+k^2}-\frac{1}{2},
    \end{equation}

    \begin{equation}
    \label{tr3}
        \omega_3=\sqrt{\frac{1}{4}+k^2}+\frac{1}{2}.
    \end{equation}
At $k\to 0$  $\omega_2\sim k^2$, i.e. it is a gapless mode arising
from the degeneracy of  precession with respect to $\alpha_0$.
Another mode has a gap $\omega_3=\omega_L$ at $k\to 0$, it passes
into nutations. In the coordinates $\psi,\vartheta,\eta,\zeta$ the
following right eigenvectors correspond to the frequencies
$\pm\omega_2$:
$$
\textbf{e}^{\pm}_2=\left(
\begin{array}{c}
1\\
\mp i\omega_3\\
\pm i\\
\omega_2
\end{array}
\right), \eqno \stepcounter{equation}(\arabic{equation})
$$
and to the frequencies $\pm\omega_3$:
$$
\textbf{e}^{\pm}_3=\left(
\begin{array}{c}
1\\
\mp i\omega_2\\
\mp i\\
-\omega_3
\end{array}
\right). \eqno \stepcounter{equation}(\arabic{equation})
$$
Left eigenvectors are correspondingly
\begin{equation}
    \label{covar3}
\textbf{f}^{\pm}_2=\frac{1}{2\omega_{23}}\left(\omega_3; \mp i;
\pm i\omega_2; 1\right),
\end{equation}
and
\begin{equation}
    \label{covar4}
\textbf{f}^{\pm}_3=-\frac{1}{2\omega_{23}}\left(-\omega_2; \pm i;
\pm i\omega_3; 1\right),
\end{equation}
where
 $\omega_{23}=\omega_2+\omega_3$.

 {\bf 3}. Time-dependent corrections to  the Hamiltonian (\ref{h_1}) can
 provide creation and mutual transformation of excitations.
 At $\textbf{l}\|\textbf{H}_0$ the dipole energy does not contain time-dependent terms and the gradient
 energy in the first approximation on $\mu$ contains oscillating
 terms (\ref{F_oscill1}),(\ref{F_oscill2}).
 Taking the oscillating terms into account one can
write equations of motion for deviations
$\varepsilon,\sigma,\nu,\vartheta,\zeta,\eta$ combined in a
six-component vector-column in a form:

    \begin{equation}
          \label{V_equation}
          \frac{d\mathbf{X}}{dt}=\left(\hat{M}_0+\hat{V}(t)\right)\mathbf{X},%\frac{d\mathbf{X}}{dt}=\hat{M}_0\mathbf{X}
    \end{equation}
where all time-dependent terms are collected in $\hat{V}(t)$. The
sum of (\ref{F_oscill1}) and (\ref{F_oscill2}) yields
$\hat{V}(t)=\sum_{n=1,2}\left[\hat{W}_n\exp(-in\omega_pt)+\hat{W}^*_n\exp(in\omega_pt)\right]$.
Following procedure of time-dependent perturbation theory let us
seek for a solution of the system (\ref{V_equation}) in a form of
expansion in eigenvectors of matrix $\hat{M}_0$
\begin{eqnarray}
         \label{zero}
         \textbf{X}(\textbf{r},t)=\sum_{j,\textbf{k}} \{ a_{j\textbf{k}}^+(t)\textbf{e}_{j\textbf{k}}^+\exp(i\textbf{k}\textbf{r}-i\omega_jt )+
         a_{j\textbf{k}}^-(t)\textbf{e}_{j\textbf{k}}^-\exp(i\textbf{k}\textbf{r}+i\omega_jt) \}
    \end{eqnarray}
Eigenvectors $\textbf{e}_{j\textbf{k}}^{\pm}$ are the mentioned
above $\textbf{e}^{\pm}_1, \textbf{e}^{\pm}_2, \textbf{e}^{\pm}_3$
supplemented with zeroes up to six component. Substitution of
(\ref{zero}) into (\ref{V_equation}) and separation of equations
by $\textbf{k}$ yields:
\begin{eqnarray}
         \label{two}
         \sum_{j} \{ \dot{a}_{j\textbf{k}}^+\textbf{e}_{j\textbf{k}}^+\exp(-i\omega_jt )+
         \dot{a}_{j\textbf{k}}^-\textbf{e}_{j\textbf{k}}^-\exp(i\omega_jt) \}=\nonumber\\\sum_{j,n}(\hat{W}_n\exp(-in\omega_pt)+\hat{W}^*_n\exp(in\omega_pt))
         \{ a_{j\textbf{k}}^+\textbf{e}_{j\textbf{k}}^+\exp(-i\omega_jt )+
         a_{j\textbf{k}}^-\textbf{e}_{j\textbf{k}}^-\exp(i\omega_jt)\}
    \end{eqnarray}
Multiplying two sides of equation (\ref{two}) by
$\textbf{f}^+_l\exp(i\omega_lt)$ and omitting common index
$\textbf{k}$ one has:
\begin{eqnarray}
         \label{three}
 \dot{a}_l^+=\sum_{j,n}
         \{a_j^-(\textbf{f}^+_l,\hat{W}_n\textbf{e}_j^-)\exp[i(\omega_l+\omega_j-n\omega_p)t] +\nonumber\\
         a_j^+(\textbf{f}^+_l,\hat{W}^*_n\textbf{e}_j^+)\exp[i(\omega_l-\omega_j+n\omega_p)t]\}.
    \end{eqnarray}
Application of the procedure assumes
 that coefficients $a_{j\textbf{k}}^{\pm}$ are weakly varying at
 time $\sim 1/\omega_e$. Averaging of equation (\ref{three}) shows that nontrivial
 correlations between different $a_{j\textbf{k}}^{\pm}$ arise
 only nearby the resonances $\omega_j(\textbf{k})-\omega_l(\textbf{k})=n\omega_p$ and
  $\omega_l(\textbf{k})+\omega_j(-\textbf{k})=n\omega_p$.
 The second resonance corresponds to creation of quasiparticles from
 "vacuum". In accordance with equality
 $\omega_j(\textbf{k})=\omega_j(-\textbf{k})$ the sign of one of the
 momenta is changed.
 In what follows it is assumed that the resonance
 condition is fulfilled exactly then the obtained increment of
 instability is a maximum one.
 If at given $\textbf{k}$ the resonance condition is fulfilled only
 for two states $l$ and $j$, then
 \begin{eqnarray}
         \label{res1}
 \dot{a}_l^+=(\textbf{f}^+_l,\hat{W}\textbf{e}_j^-)a_j^-.
    \end{eqnarray}
The same argument for  $a_j^-$ yields:
\begin{eqnarray}
         \label{res2}
 \dot{a}_j^-=(\textbf{f}^-_j,\hat{W}^*\textbf{e}_l^+)a_l^+.
    \end{eqnarray}
The system of equations (\ref{res1}),(\ref{res2}) has solutions
$\sim\exp(\pm\lambda t)$, where $\lambda$ is determined by
$\lambda^2=(\textbf{f}^+_j,\hat{W}\textbf{e}_l^-)(\textbf{f}^+_l,\hat{W}\textbf{e}_j^-)^*$.
Thus the problem of determination of the increment of instability
reduces to calculation of the elements of matrix
$(\textbf{f}^+_j,\hat{W}\textbf{e}_l^-)$ between the states
satisfying the resonance conditions. Particularly, if $l=j$, then
$2\omega_l(\textbf{k})=n\omega_p$ and
$\lambda^2=|(\textbf{f}^+_j,\hat{W}\textbf{e}_l^-)|^2$. This is the
simplest case of \emph{parametric resonance} \cite{Landau}. Explicit
expressions for $\hat{W}_1$ and $\hat{W}_2$ are found using
equalities (\ref{F_oscill1}) and (\ref{F_oscill2}). There are four
non-zero elements of matrix $\hat{W}_1$:
$(\hat{W}_1)_{\sigma\nu}$=$(\hat{W}_1)_{\vartheta\varepsilon}$=$i\mu
k_yk_z$ and
$(\hat{W}_1)_{\sigma\eta}$=$(\hat{W}_1)_{\zeta\varepsilon}$=$\mu
k_yk_z$, and there are another four finite elements of matrix
$\hat{W}_2$:
$(\hat{W}_2)_{\vartheta\nu}$=-$(\hat{W}_2)_{\zeta\eta}$=$\mu
k_y^2/2$ and
$(\hat{W}_2)_{\vartheta\eta}$=$(\hat{W}_2)_{\zeta\nu}$=$i\mu
k_y^2/2$. Here $k_y, k_z$ are the components of the wave vector
satisfying the resonance conditions. Direct verification shows that
the matrix element $(\textbf{f}^+_j,\hat{W}_1\textbf{e}_l^-)$ is not
equal to zero only if  $j=1, l=3$. The resonance condition
$\omega_1(\textbf{k})+\omega_3(-\textbf{k})=\omega_L $ is fulfilled
only at $k=0$, but at this value of $k$ the matrix element turns
into zero due to the factor $\ k_yk_z$. For the matrix
$(\textbf{f}^+_j,\hat{W}_2\textbf{e}_l^-)$ non-zero element
corresponds to $j=l=3$. In this case the resonance condition
$\omega_3(\textbf{k})+\omega_3(-\textbf{k})=2\omega_L$ is also
fulfilled only at $k=0$ and the corresponding matrix element is
equal to zero due to the factor $\ k_y^2$.  Thus if the dipole
energy is neglected then the anisotropy of spin wave velocities does
not provide the coupling between precession and spin waves.

{\bf 4}. In order to take the dipole energy into account
 it is need to add to the RHS of
 equation (\ref{V_equation}) the matrix of dipole torque $\hat{N}_D$
 and to find new energies of excitations and the corresponding
 eigenvectors.
 The added terms in the
 equation (\ref{V_equation}) are small in comparison with the
 elements of the matrix $\hat{M}_0$ as $(\Omega_B/\omega_L)^2$,
 which is in typical experimental conditions of the order of $\sim
 10^{-1}-10^{-2}$. In the first order approximation on the mentioned
 parameter the elements of matrix
 $(\textbf{f}^+_j,\hat{W}_n\textbf{e}_l^-)$ are given by the
 expressions:
 \begin{eqnarray}
         \label{pert}
 (\textbf{f}^+_j,\hat{W}_n\textbf{e}_l^-)=i\sum_{m\neq j+}
         \frac{1}{\omega_{j+}-\omega_m}(\textbf{f}^+_j,\hat{N}_D\textbf{e}_m)
         (\textbf{f}_m,\hat{W}_n\textbf{e}_l^-) +\nonumber\\
         i\sum_{m\neq l-}
         \frac{1}{\omega_{l-}-\omega_m}(\textbf{f}^+_j,\hat{W}_n\textbf{e}_m)
         (\textbf{f}_m,\hat{N}_D\textbf{e}_l^-).
    \end{eqnarray}

Non-zero elements are obtained at the following resonances:
\begin{eqnarray}
\omega_1(k)+\omega_1(-k)=\omega_p,~~ k=1/2,~~ \beta\leq\theta_0,\\
\omega_1(k)+\omega_2(-k)=\omega_p,~~ k=2/3,~~~~~~~~~~~~\\
\omega_1(k)+\omega_3(-k)=2\omega_p,~~ k=2/3,~~~~~~~~~~\\
\omega_2(k)+\omega_3(-k)=2\omega_p,~~ k=\sqrt{3}/2.~~~~~~~~
  \end{eqnarray}
The corresponding increments are of the order of
$\mu\Omega_B^2/\omega_L$. At $k$=1/2
$$
\lambda_{(11)}= \left\{
\begin{array}{cc}
\displaystyle \mu\sin(2\delta)\frac{\sqrt{(1-\cos(\beta))(1+4\cos(\beta))}}{5}\frac{\Omega_B^2}{\omega_L},&\beta\leq\theta_0,\\
0&\beta>\theta_0,
\end{array}
\right.
 \eqno \stepcounter{equation}(\arabic{equation}) \label{inc1}
$$
where $\delta$ is the angle between the direction of the wave
vector $\textbf{k}$ and the direction of the magnetic field. At
$k$=2/3 there are two resonances and one needs to consider the
system of equations for thee amplitudes in order to find the
increment. In this case the increment is given by the expression
\begin{equation}
\lambda_{(12,13)}=\sqrt{\lambda_{(12)}^2+\lambda_{(13)}^2},
\end{equation}
where
$$
\lambda_{(12)}= \left\{
\begin{array}{cc}
\displaystyle \mu\sin(2\delta)\frac{\sqrt{10}(1-\cos(\beta))}{25}\frac{\Omega_B^2}{\omega_L},&\beta\leq\theta_0,\\
\displaystyle
\mu\sin(2\delta)\frac{|4\cos^2(\beta)+31\cos(\beta)+15|}{15\sqrt{10}}\frac{\Omega_B^2}{\omega_L},&\beta>\theta_0,
\end{array}
\right.
 \eqno \stepcounter{equation}(\arabic{equation}) \label{inc21}
$$
$$
\lambda_{(13)}= \left\{
\begin{array}{cc}
\displaystyle \mu\sin^2(\delta)\frac{\sqrt{10}\sqrt{(1-\cos(\beta))(1+4\cos(\beta))}}{25}\frac{\Omega_B^2}{\omega_L},&\beta\leq\theta_0,\\
\displaystyle
-\mu\sin^2(\delta)\frac{\sqrt{10}\sin(\beta)(1+4\cos(\beta))}{225}\frac{\Omega_B^2}{\omega_L},&\beta>\theta_0,
\end{array}
\right.
 \eqno \stepcounter{equation}(\arabic{equation}) \label{inc22}
$$
The increment for the resonance at $k=\sqrt{3}/2$ is
$$
\lambda_{(23)}= \left\{
\begin{array}{cc}
\displaystyle \mu\sin^2(\delta)\frac{3(1-\cos(\beta))}{20}\frac{\Omega_B^2}{\omega_L},&\beta\leq\theta_0,\\
\displaystyle
\mu\sin^2(\delta)\frac{|4\cos^2(\beta)+31\cos(\beta)+15|}{40}\frac{\Omega_B^2}{\omega_L},&\beta>\theta_0.
\end{array}
\right.
 \eqno \stepcounter{equation}(\arabic{equation}) \label{inc22}
$$
The maximum increment for each value of $\beta$  can be written as
\begin{equation}
\label{a}
\lambda_{max}(\beta)=a(\beta)\mu\frac{\Omega^2_B}{\omega_L},
\end{equation}
where the dependence of the coefficient $a$ on tipping angle is
shown at Fig.1.
\begin{figure}
\begin{center}
\includegraphics[%
  width=0.55\linewidth,
  keepaspectratio]{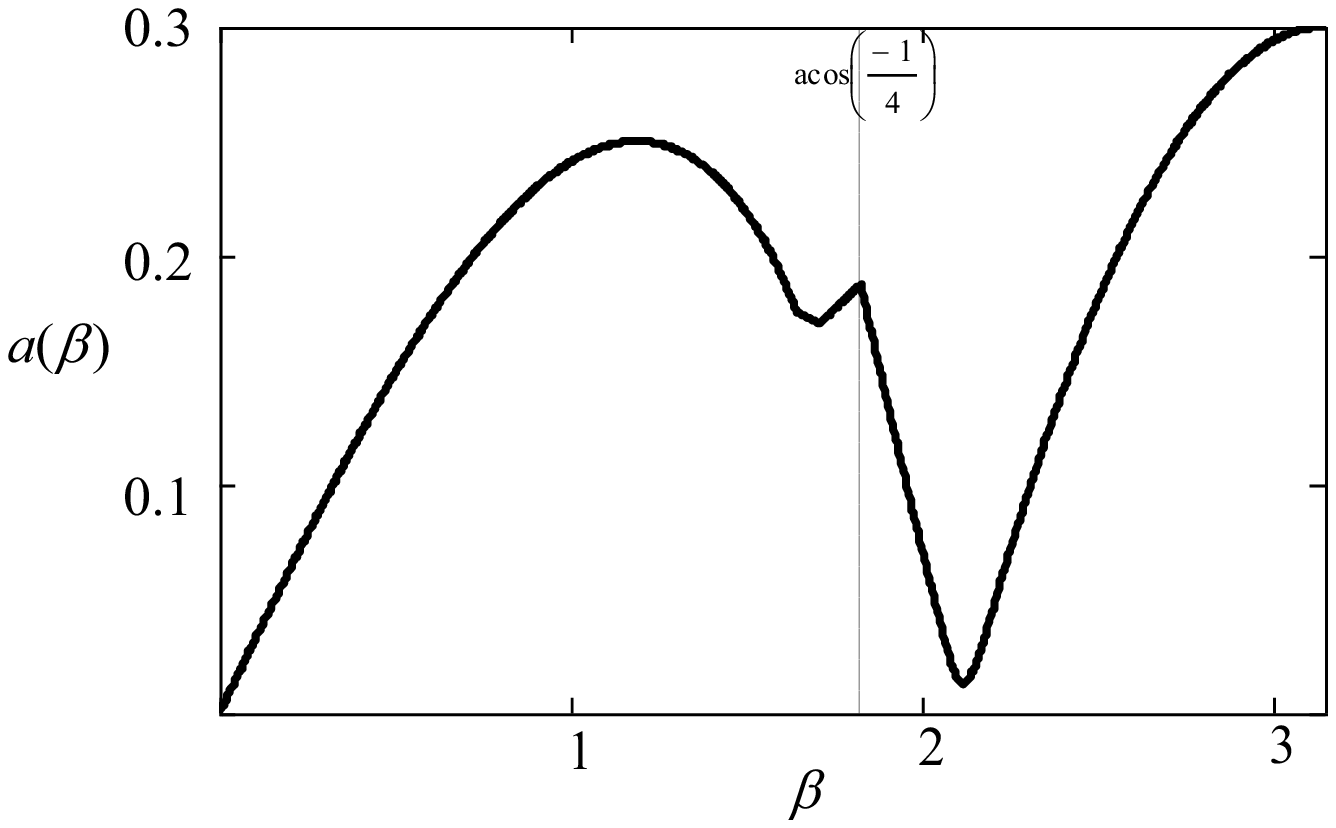} %[width=3in,height=2in]
\end{center}
\caption ~ Dependence of the coefficient $a$ in formule (\ref{a}) on
tipping angle  $\beta$\label{frame1}
\end{figure}

At finite temperatures the damping of spin waves has to be taken
into account. The instability sets up if the increment of growth
of waves, which satisfy the resonance condition, exceeds the
decrement of damping. As before \cite{sf} for estimation of the
temperature of catastrophic relaxation $T_{cat}$ it will be
assumed here that the principal mechanism of dissipation is spin
diffusion. The minimum temperature, down to which the precession
is stable, is found from the equation:
\begin{equation}
\label{dif} \frac{D(T)k^2}{2}=\lambda_{max},
\end{equation}
where $D(T)$ is the coefficient of spin diffusion. At temperatures
in question $T\leq 0.4T_c$, the increment weakly depends on
temperature and its value can be taken at  $T=0$. The LHS of
(\ref{dif}) strongly depends on temperature owing to spin diffusion
which behaves as $D(T)\sim\sqrt{T/\Delta}\exp(-\Delta/T)$ at
$T\rightarrow 0$.

The obtained increments originate from the coupling between
precession and spin waves in bulk helium. In Ref. \cite{blv1,blv2}
the increment arising from enhancement of the coupling in regions
adjacent to the walls was found. Because of boundary conditions
precession on the walls goes on in the configuration different from
the Leggett one. In this case oscillating terms in the dipole energy
appear without taking the anisotropy of spin waves velocities into
account. The local coupling in regions adjacent to the walls is on
the order of $1/\mu$ greater than that obtained for the bulk helium.
In Ref.\cite{blv1,blv2} the result of calculation of increment of
growth and $T_{cat}$ for the angle $\beta=90^{\circ}$ is
represented. The bulk contribution to the increment for the same
conditions estimated with the use of the above formulae gives
approximately the same value as the surface contribution. It is
impossible to separate surface and bulk contributions to the
increment by their dependence on magnetic field because both
contributions are proportional to $\Omega_B^2/\omega_L$. However, it
has to be mentioned, that the surface contribution depends on ratio
of volume adjacent to the surfaces to the total volume of helium.
For angles $\beta<104^{\circ}$ there is no characteristic length
whereon penetrates the perturbation effect of the walls on the
precession. Thus the regions adjacent to the walls occupies
considerable part of the volume. The most part of the data about
catastrophic relaxation is obtained in experiments with the
homogeneously precessing domain. In this case magnetization
precesses with angles slightly above $\theta_0\approx104^{\circ}$
and the frequency of precession is shifted from the Larmor
frequency. Then the effect of the walls is limited by the "coherence
length"$~$ $\xi=c_{\perp}/\sqrt{\omega_p(\omega_p-\omega_L)}$
\cite{fm2,tim}. For the typical experimental conditions in the most
part of the precessing domain $\xi\sim 10^{-2}$cm and the regions
adjacent to the walls occupies only a small part of the total
volume. The increment is determined by the bulk resonance
$\omega_2(k)+\omega_3(-k)=2\omega_p$. The available experimental
data for diffusion coefficient \cite{Experiment} pertains to the
temperatures $T> 0,4 T_c$. In the experiments \cite{nyeki} at
pressure $P\simeq 20$ bar and magnetic field $H\simeq 142$ Oe
$T_{cat}\simeq 0.42 T_c$. After substitution of the diffusion
coefficient at this temperature $D= 0.04$ cm$^2/c$ and the values of
other parameters $\Omega= 2\pi \cdot250$ kHz,$c_{\parallel}=
1.6\cdot 10^3$ cm/c, $\mu=1/4$ into the formula (\ref{dif}) we
obtain for the LHS of equation (\ref{dif}) the value
5$\cdot$10$^4$1/c, and for the RHS -- 4$\cdot$10$^4$1/c. Taking into
account that parameters $\mu$ and $\Omega_B^2/\omega_L^2$ are not
too small the obtained agreement can be regarded as a satisfactory
for this example.

{\bf 5}. In conclusion the Suhl instability limits from below the
interval of temperatures where coherent precession in $^{3}He-B$ can
exist. The obtained here low temperature limit of stability of
precession is caused by the interaction of precession with spin
waves in the bulk helium.

Lowering of the limit temperature can be achieved by using higher
magnetic fields as it was demonstrated in experiments \cite{Lee}.
Such tendency agrees with formula (\ref{dif}). On the one hand
increasing of magnetic field decreases the increment of
instability and on the other hand it increases spin waves damping.
However, due to the exponential dependence of the coefficient of
diffusion on temperature the effect of magnetic field on $T_{cat}$
becomes weaker when temperature decreases. Detailed comparison
with the results of \cite{Lee} will be done in a full-length
publication.

We thank to V.V. Dmitriev for useful discussion. This research was
supported by RFBR and Ministry of Education and Science of Russian
Federation.


\begin{thebibliography}{80}
\bibitem{nyeki}  Yu.M. Bunkov, V.V. Dmitriev, Yu.M. Mukharsky et al.,
{\it Europhysics Lett.} {\bf 8}, 645 (1989).
\bibitem{Bunkov_precession} Yu.M. Bunkov, V.V. Dmitriev, J. Nyeki et al., Physica
B \textbf{165}, 675 (1990).
\bibitem{sf}  E.V.Surovtsev, I.A. Fomin, Pis'ma ZheTF {\textbf{83}}, 479
(2006).
\bibitem{suhl} H. Suhl, J. Phys. Chem. Solids, {\bf 1}, 209 (1957).
\bibitem{Lee} D.A. Geller and D.M. Lee, Phys. Rev.
Lett. \textbf{85}, 1032 (2000).
\bibitem{blv1} Yu.M. Bunkov, V.S. Lvov, G.E. Volovik, Pis'ma v
ZhETF \textbf{83}, 624 (2006)
\bibitem{blv2} Yu.M. Bunkov, V.S. Lvov, G.E. Volovik, Pis'ma v
ZhETF \textbf{84}, 349 (2006)
\bibitem{Fomin1}I. A. Fomin, Zh. Exp. Teor. Fiz. {\bf 84}, 2109~(1983) [Sov. Phys. JETP {\bf 57,} 1227~(1983)].
\bibitem{dm_osc}  Yu.M. Bunkov, V.V. Dmitriev, Yu.M. Mukharsky, Pis'ma ZheTF \textbf{43}, 131 (1986)
\bibitem{Landau} L. D. Landau and E. M. Lifshitz, {\it Course of Theoretical
Physics,} Vol. 1: {\it Mechanics}, 4th ed. (Nauka, Moscow, 1988,
Pergamon, Oxford, 1989)
\bibitem{fm2} I. A. Fomin, Zh. Exp. Teor. Fiz. {\bf 94}, 112~(1988).
\bibitem{tim} Yu.M. Bunkov, O.D. Timofeevskaya, G.E. Volovik, Phys. Rev. Lett. {\bf
73}, 1817 (1994).
\bibitem{Experiment} Yu.M. Bunkov, V.V. Dmitriev, A.V. Markelov et al., Phys. Rev. Lett. {\bf
65}, 867 (1990).


\end{thebibliography}
\end{document}